\documentclass[pra]{revtex4-2}
\usepackage{eurosym}
\usepackage{amssymb}
\usepackage{graphicx}
\usepackage{verbatim}
\usepackage[normalem]{ulem}
\usepackage{color}
\usepackage{amsmath}

\begin{document}

\title{Angular-momentum modes in a bosonic condensate trapped in the
inverse-square potential}
\author{Hidetsugu Sakaguchi$^{1}$ and Boris A. Malomed$^{2,3}$}
\affiliation{$^{1}$Department of Applied Science for Electronics and Materials,
Interdisciplinary Graduate School of Engineering Sciences, Kyushu
University, Kasuga, Fukuoka 816-8580, Japan\\
$^{2}$Department of Physical Electronics, School of Electrical Engineering,
Faculty of Engineering, Tel Aviv University, Tel Aviv 69978, Israel\\
$^{3}$Instituto de Alta Investigaci\'{o}n, Universidad de Tarapac\'{a},
Casilla 7D, Arica, Chile\footnote{Sabbatical address}}

\begin{abstract}
In the mean-field approximation, the well-known effect of the critical
quantum collapse in a 3D gas of particles pulled to the center by potential $%
U(r)=-U_{0}/\left( 2r^{2}\right) $ is suppressed by repulsive inter-particle
interactions, which create the otherwise non-existing \textit{s}-wave ground
state. Here, we address excited bound states carrying angular momentum, with
the orbital and magnetic quantum numbers, $l$ and $m$. They exist above a
threshold value of the potential's strength, $U_{0}>l(l+1)$. The sectoral,
tesseral, and zonal modes, which correspond to $m=l$, $0<m<l$, and $m=0$,
respectively, are found in an approximate analytical form for relatively
small values of $U_{0}-l(l+1)$. Explicit results are produced for the
\textit{p}- and \textit{d}-wave states, with $l=1$ and $2$, respectively. In
the general form, the bound states are obtained numerically, confirming the
accuracy of the analytical approximation.
\end{abstract}

\maketitle

\section{Introduction and the model}

It is well known that attractive potential $U(r)=-U_{0}/\left( 2r^{2}\right)
$, with $U_{0}>0$, plays a special role in quantum mechanics \cite{LL}.
Indeed, unlike the classical equation of motion in the same potential, which
gives rise to equivalent solutions for all values of $U_{0}$, due to the
scaling invariance of the equation, $U_{0}$ cannot be scaled out in the
corresponding Schr\"{o}dinger equation, written in the normalized form,%
\begin{equation}
i\psi _{t}=-\frac{1}{2}\left( \nabla ^{2}+\frac{U_{0}}{r^{2}}\right) \psi ,
\label{Schr}
\end{equation}%
where $U_{0}$ is an irreducible parameter. It is known that the
three-dimensional (3D) equation (\ref{Schr}) gives rise to the spherically
symmetric ground state (GS) if the strength of the pull to the center does
not exceed a critical value,%
\begin{equation}
U_{0}<\left( U_{0}\right) _{\mathrm{cr}}=1/4,  \label{1/4}
\end{equation}%
while at $U_{0}>1/4$ Eq. (\ref{Schr}) does not maintain the GS, leading,
instead, to the \textit{quantum collapse} of wave function $\psi $ (called
\textquotedblleft fall onto the center" of the quantum particle in the book
by Landau and Lifshitz \cite{LL}). Exact solutions of Eq. (\ref{Schr}) for
the developing collapse, as well as the respective solutions for the
classical equation of motion, were recently analyzed in Ref. \cite{Trib}.
Such a difference between the classical and quantum behavior under the
action of the same potential is considered as the \textit{quantum anomaly},
alias \textquotedblleft dimensional transmutation" \cite{anomaly,anomaly2}
(quantum anomalies for particles trapped in an external potential are also
known in the 2D geometry \cite{Olshanii}).

A physical realization of the Schr\"{o}dinger equation (\ref{Schr}) is
possible for a particle with a permanent electric dipole moment (such as a
small polar molecule) pulled to the electric charge fixed at the center. If
the particle minimizes its energy by aligning the polarization of the dipole
moment with the local electric field, the attraction potential takes
precisely the form adopted in Eq. (\ref{Schr}) \cite{HS1}. In this case, for
the particle with the mass $\sim 100$ proton masses and the elementary
electric charge set at the center, the critical value (\ref{1/4})
corresponds to a very small dipole moment, $d_{\mathrm{cr}}\sim 10^{-6}$
Debye. Therefore, the overcritical case of $U_{0}>1/4$ is relevant in the
actual physical context.

On the other hand, recently new techniques have been elaborated for the
creation of efficient optical trapping potentials for ultracold atoms, the
commonly known ones being red- and blue-shifted optical lattices \cite%
{Portugal,Morsch,Lewenstein}. More sophisticated potentials can be
\textquotedblleft painted" in space by fast moving laser beams \cite%
{Henderson}. Recently, a technique based on the use of digital micromirror
devices was elaborated for the creation of various traps for 2D
Bose-Einstein condensates (BECs) \cite{DMD}. In particular, it may be used
for inducing anharmonic potentials $\sim r^{k}$, with $k\geq 2$, that were
explored as a setting for the control of 2D vortex dynamics in BEC \cite%
{Fetter}. In this connection, it is relevant to mention that the
pull-to-the-center potential $\sim -1/r^{2}$ for an atom in the 3D space can
be induced by red-shifted laser illumination provided by beams isotropically
converging towards $r=0$, resembling that igniting the inertial nuclear
fusion (with the power larger by many orders of magnitude) \cite%
{fusion,fusion2}. Indeed, the energy of the detuned dipole interaction of
the atom with the optical field of local intensity $I(r)$ is%
\begin{equation}
U_{\mathrm{dip}}(r)=-\frac{3\pi c_{0}^{2}}{2\omega _{0}^{3}}\frac{\Gamma }{%
\Delta }I(r),  \label{Udip}
\end{equation}%
where $c_{0}$ is the light speed in vacuum, $\omega _{0}$ the frequency of
the atomic optical transition, $\Delta $ its detuning from the driving
frequency, and $\Gamma $ the linewidth \cite{dip-interaction}. Then, in the
geometric-optics approximation, the substitution of
\begin{equation}
I(r)=P/\left( 4\pi r^{2}\right) ,  \label{I}
\end{equation}%
where $P$ is the integral illumination power, in Eq. (\ref{Udip}), produces
the potential sought for. Then, an estimate for $\omega $, corresponding for
the wavelength of the laser beams $\lambda \sim 1$ $\mathrm{\mu }$m and the
lightest atom (hydrogen), demonstrates that the critical value (\ref{1/4})
corresponds to extremely small powers, $P_{\mathrm{cr}}\sim 10^{-7}$ W,
hence the overcritical case is fully relevant for the experimental
implementation.

A possible solution of the quantum-collapse problem, i.e., restoration of
the missing GS in the case of $U_{0}>1/4$, was proposed in terms of the
secondary quantization, replacing the Schr\"{o}dinger equation (\ref{Schr})
by the corresponding linear quantum-field theory \cite{anomaly,anomaly2}.
However, the solution does not predict the size of the so introduced GS.
Instead, the renormalization-group technique, on which the field-theory
formulation is based, postulates the existence of a GS with an \emph{%
arbitrary} spatial size, in terms of which all other spatial scales are
measured in the framework of the theory.

A completely different, many-body, solution of the collapse problem was
proposed in Ref. \cite{HS1}. Instead of addressing the single particle, it
deals with an ultracold gas of particles pulled to the center, in the form
of BEC. In particular, ultracold gases of polar molecules such as LiCs \cite%
{LiCs} and KRb \cite{KRb} are available to the experiment, and it was
demonstrated too that a fixed electric charge (ion) can be embedded in BEC
and held at a fixed position by means of an optical trapping scheme \cite%
{ion}.

The suppression of the quantum collapse in this setting is secured by
repulsive collisional interactions between the particles. The analysis was
performed in the framework of the mean-field (MF) approximation \cite{BEC},
using the Gross-Pitaevskii equation (GPE) with the same potential as in Eq. (%
\ref{Schr}), In the properly scaled form, the GPE id written as
\begin{equation}
i\psi _{t}=-\frac{1}{2}\left( \nabla ^{2}+\frac{U_{0}}{r^{2}}\right) \psi
+\left\vert \psi \right\vert ^{2}\psi ,  \label{GPE}
\end{equation}%
where the cubic term in Eq. (\ref{GPE}) represents the self-repulsion in the
gas, in the framework of the MF approximation. The coefficient in front of
this term is set equal to $1$ by means of normalization. Dipole-dipole
interactions between the particles in the gas were also taken into account,
applying another version of the MF approximation. It considers the
interaction of a given dipole with the collective electric potential created
by all other dipoles, and amounts to renormalization of coefficient $U_{0}$.

In addition to the irreducible coefficient $U_{0}$, another control
parameter of the MF theory is the norm,%
\begin{equation}
N=\int \int \int \left\vert \psi \left( x,y,z\right) \right\vert ^{2}dxdydz.
\label{N}
\end{equation}%
Alternatively, it is possible to fix that the norm is $1$, while the cubic
term in Eq. (\ref{GPE}) will acquire coefficient $N$ as a free parameter.

For a mixture of two species of particles, a two-component extension of Eq. (%
\ref{GPE}) was introduced in Ref. \cite{HS3}. The respective system of
nonlinearly coupled GPEs also provides the suppression of the quantum
collapse and creation of the GS \cite{HS3}.

Solutions of Eq. (\ref{GPE}) for stationary states with chemical potential $%
\mu <0$, in the form of
\begin{equation}
\psi =e^{-i\mu t}u\left( x,y,z\right) ,  \label{psi-u}
\end{equation}%
with steady-state wave function $u$ satisfying the equation%
\begin{equation}
\mu u=-\frac{1}{2}\left( \nabla ^{2}+\frac{U_{0}}{r^{2}}\right) u+\left\vert
u\right\vert ^{2}u.  \label{u}
\end{equation}%
These states are characterized by their norm (\ref{N}) and angular momentum,%
\begin{equation}
M=i\int \int \int \psi ^{\ast }\left( x,y,z\right) \left( \mathbf{r}\times
\nabla \right) \psi \left( x,y,z\right) dxdydz,  \label{M}
\end{equation}%
where $\ast $ stands for the complex conjugate. $N$ and $M$ are dynamical
invariants of GPE (\ref{GPE}). It also conserves the Hamiltonian,

\begin{equation}
E=\frac{1}{2}\int \int \int \left[ \left\vert \nabla \psi \right\vert ^{2}-%
\frac{U_{0}}{r^{2}}\left\vert \psi \right\vert ^{2}+\left\vert \psi
^{4}\right\vert \right] dxdydz.  \label{E}
\end{equation}

The linearized version of Eq. (\ref{GPE}) produces exact, although
unnormalizable (with diverging norm (\ref{N})), pairs of eigenmodes with $%
\mu =0$ and orbital and magnetic quantum numbers $l,m$ \cite{HS1},%
\begin{equation}
u=u_{0}r^{-\sigma _{\pm }}\mathrm{Y}_{lm}\left( \theta ,\varphi \right)
,\sigma _{\pm }=\frac{1}{2}\pm \sqrt{\frac{1}{4}-U_{l}},  \label{Y}
\end{equation}%
which are written in spherical coordinates $\left( r,\theta ,\varphi \right)
$. Here $u_{0}$ is an arbitrary amplitude, $\mathrm{Y}_{lm}$ are the
spherical harmonics, and%
\begin{equation}
U_{l}\equiv U_{0}-l\left( l+1\right) .  \label{-l(l+1)}
\end{equation}%
Solution (\ref{Y}) implies that these eigenstates exists at $U_{l}<1/4$, cf.
Eq. (\ref{1/4}), while the quantum collapse occurs at $U_{l}>1/4$.

The suppression of the collapse and recreation of the GS (with a
normalizable wave function, unlike the unnormalizable one (\ref{Y})) in the
framework of the full nonlinear GPE (\ref{GPE}) was demonstrated for the
spherically isotropic state, corresponding to $l=m=0$ in expression (\ref{Y}%
) (i.e., the \textit{s}-wave orbital, in terms of atomic physics \cite{LL}).
To this end, Eq. (\ref{u}) is converted, by substitution

\begin{equation}
u\left( r\right) =r^{-1}v(r)  \label{uv}
\end{equation}%
with real function $v(r)$, into a radial equation%
\begin{equation}
\mu v=-\frac{1}{2}\left( \frac{d^{2}}{dr^{2}}+\frac{U_{0}}{r^{2}}\right) v+%
\frac{v^{3}}{r^{2}}.  \label{chi}
\end{equation}%
An asymptotic solution to Eq. (\ref{chi}) at $r\rightarrow 0$ is
\begin{equation}
v(r)=\sqrt{U_{0}/2}+\mathrm{const}\cdot r^{s/2},~s\equiv 1+\sqrt{1+8U_{0}}
\label{r=0}
\end{equation}%
(here, $\mathrm{const}$ is not determined by the asymptotic expansion). At $%
r\rightarrow \infty $, the bound-state wave function with $\mu <0$ decays
exponentially, $v\sim \exp \left( -\sqrt{-2\mu }r\right) $. A global
approximation for the GS can be constructed as an interpolation, juxtaposing
the asymptotic forms which are valid at $r\rightarrow 0$ and $r\rightarrow
\infty $:{\
\begin{equation}
\left( \psi _{\mathrm{GS}}(r,t)\right) _{\mathrm{interpol}}\approx \sqrt{%
\frac{U_{0}}{2}}e^{-i\mu t}r^{-1}e^{-\sqrt{-2\mu }r}.~  \label{inter}
\end{equation}%
}

Solutions with $\mu >0$ correspond to delocalized states with the following
asymptotic form at $r\rightarrow \infty $:%
\begin{equation}
u(r)\approx \sqrt{\mu }+\frac{U_{0}}{4\sqrt{\mu }}r^{-2}+\frac{U_{0}}{8\mu
^{3/2}}\left( 1-\frac{U_{0}}{4}\right) r^{-4}+...~.  \label{mu > 0}
\end{equation}%
These states are not considered here in detail.

{Singularity }$\sim r^{-1}$ {of wave function (\ref{inter}) at }$%
r\rightarrow 0$ is acceptable, as the corresponding 3D norm integral (\ref{N}%
) converges at $r\rightarrow 0$. In particular, the approximate wave
function (\ref{inter}) gives rise to the following relation between the
chemical potential and norm,%
\begin{equation}
\mu _{\mathrm{GS}}\approx -\frac{1}{2}\left( \frac{\pi U_{0}}{N_{\mathrm{%
interpol}}}\right) ^{2}.  \label{mu}
\end{equation}%
Actually, scaling $\mu \sim -1/N^{2}$, which is produced by Eq. (\ref{mu}),
is an exact property\emph{\ }of solutions to Eq. (\ref{u}), irrespective of
validity of the approximation. In the limit of $\mu \rightarrow -0$, Eq. (%
\ref{inter}) produces an exact solution of Eq. (\ref{GPE}), $\psi _{\mu
=0}(r)=\sqrt{U_{0}/2}r^{-1}$, with a divergent norm.

The numerical solution of Eq. (\ref{u}) corroborates the conclusion
following from Eqs. (\ref{r=0}) -- (\ref{mu}): GPE\ (\ref{GPE}) does not
give rise to the collapse at $U_{0}>1/4$, and maintains the well-defined GS
at all values of $U_{0}$. For the setting with the elementary electric
charge fixed at the center, and particles with the mass $\sim 100$ proton
masses and electric dipole moment $\sim 1$ Debye, the radius of the newly
predicted GS, which replaces the collapsing regime in the gas of $\sim
10^{5} $ particles, is $r_{\mathrm{GS}}\simeq 3$ $\mathrm{\mu }$m \cite{HS1}%
. Note that the MF approximation is relevant for states characterized by
such a length scale.

As concerns the implementation of the model with the pull-to-the center
optical potential provided by Eqs. (\ref{Udip}) and (\ref{I}), an estimate
demonstrates that the same order of magnitude of the GS radius is expected
for the gas of $\sim 10^{6}$ atoms under the action of the isotropically
converging illumination with power $P\sim 10^{-3}$ W.

Beyond the framework of the MF approximation, the many-body theory
demonstrates that the predicted state persists as a meta-stable one,
separated by a tall potential barrier from the collapsing state (the
many-body analysis does not allow full suppression of the collapse) \cite%
{GEA}. For the number of particles $\gtrsim 100$, the barrier is practically
impenetrable, hence the MF-predicted solution becomes an effective GS.

Relation (\ref{mu}) satisfies the \textit{anti-Vakhitov-Kolokolov criterion}%
, $d\mu /dN>0$, which is a necessary condition for stability of families of
bound states supported by a self-repulsive nonlinearity \cite{anti} (the
Vakhitov-Kolokolov criterion per se, $d\mu /dN<0$, pertains to the case of
the self-attraction \cite{VK,Berge'}). The stability of the GS created by
Eq. (\ref{GPE}) for all considered values of $U_{0}$ was corroborated by
systematic simulations of the perturbed evolution of the bound states \cite%
{HS1}.

An obviously interesting extension of the analysis outlined above, which is
the subject of the present work, is to develop it for bound states with
reduced symmetry, which carry angular momentum. Previously, states with
embedded vorticity, which is directly related to the angular momentum, were
considered in a modified setting, with the polarization of dipole moments
fixed by an external uniform electric field, directed along the $z$ axis
\cite{HS2}. In that case, the spherically symmetric pulling potential in Eq.
(\ref{GPE}) is replaced by an axially symmetric one, $-\left(
U_{0}/2r^{2}\right) \cos \theta $. Critical values of $U_{0}$ above which
the \emph{linear} axially-symmetric model gives rise to the quantum collapse
of states with magnetic quantum numbers (vorticities) $m=0$ (i.e., GS), $1$
and $2$ are, respectively, $\left( U_{0}\right) _{\mathrm{cr}}^{(m)}=1.28$, $%
7.58$, and $19.06$. For comparison, Eqs. (\ref{1/4}) and (\ref{1}) yield the
set of smaller critical values, \textit{viz}. $\left( U_{0}\right) _{\mathrm{%
cr}}^{(l)}=0.25$, $2.25$, and $6.25$ for $l=0$, $1$, and $2$, respectively.
The nonlinear model with the axisymmetric potential also suppresses the
collapse and creates stable bound states with vorticity $m$ at $U_{0}>\left(
U_{0}\right) _{\mathrm{cr}}^{(m)}$.

Vortex states were also addressed in the 2D version of the setting, whose
linearized form gives rise to the collapse at all positive values of $U_{0}$%
. In that case, the cubic self-repulsion in GPE is not sufficient to
suppress the collapse \cite{HS1}. A physically relevant alternative is
provided by the quartic term, which is produced by the Lee-Huang-Yang
effect, i.e., a correction to the MF theory induced by quantum fluctuations
\cite{LHY}. To this end, one may consider a binary BEC in which the MF
self-repulsion in each component is (almost) precisely balanced by the
inter-component attraction \cite{Petrov,AP}. The resulting amended GPE
predicts self-trapped states in the form of \textit{quantum droplets}, which
have been observed experimentally with quasi-2D \cite{droplet1,droplet2} and
full 3D \cite{droplet3,droplet4} shapes in binary BECs with the contact
interactions, as well as in dipolar BECs with long-range interactions
between atomic magnetic dipole moments \cite{Santos,Pfau}.

In the present context, the corresponding 2D GPE takes the form of Eq. (\ref%
{GPE}), with term $|\psi |^{2}\psi $ replaced by $|\psi |^{3}\psi $ \cite%
{Shamriz}, leading to the solution for the wave function with the following
asymptotic form at $r\rightarrow 0$, which replaces the collapsing solution
of the 2D linear equation:
\begin{equation}
\psi \approx \left[ \frac{1}{2}\left( U_{m}+\frac{4}{9}\right) \right]
^{1/3}e^{i\mu t-im\varphi }r^{-2/3},~U_{m}\equiv U_{0}-m^{2}  \label{2D}
\end{equation}%
(cf. expression (\ref{inter}) at $r\rightarrow 0$ and Eq. (\ref{-l(l+1)})),
where $\left( r,\varphi \right) $ are the 2D polar coordinates, and integer $%
m$ is the vorticity. Note that the 2D norm of this wave function with the
\textit{acceptable} \textit{singularity} $r^{-2/3}$ converges at $%
r\rightarrow 0$. A nontrivial problem is stability of the 2D vortex bound
states corresponding to the asymptotic form (\ref{2D}), which exist under
the condition of $U_{0}>m^{2}-4/9\equiv \left( U_{0}\right) _{\min }$.
Surprisingly, this problem admits an exact solution: the vortex states with $%
m\geq 1$ are stable provided that $U_{0}>(7/9)\left( 3m^{2}-1\right) \equiv
\left( U_{0}\right) _{\mathrm{stability}}$, the GS with $m=0$ being stable
too \cite{Shamriz}. Note that, for all $m\geq 1$, $\left( U_{0}\right) _{%
\mathrm{stability}}$ exceeds $\left( U_{0}\right) _{\min }$, hence the 2D
vortex states are unstable in the interval of $m^{2}-4/9<U_{0}<$ $%
(7/9)\left( 3m^{2}-1\right) $. The dominant instability mode is one which
leads to slow drift of the vortex' pivot off the central point, along a
spiral trajectory \cite{Shamriz}.

The objective of the present work is to construct 3D states, carrying
angular momentum, as solutions of GPE (\ref{GPE} with nonzero orbital
quantum number, $l\geq 1$, and the magnetic quantum number taking values $%
0\leq m\leq l$. Essential results for this problem can be obtained in an
analytical form, as shown below in Section 2. In particular, an exact
existence threshold is found for all nonlinear states with $l\geq 1$,
\textit{viz}.,
\begin{equation}
U_{0}\geq \left( U_{0}^{(l)}\right) _{\mathrm{thr}}=l(l+1)  \label{thr}
\end{equation}%
(it does not depend on $m$). Numerical results, which essentially
corroborate the analytical predictions, are reported in Section 3, and the
paper is concluded by Section 4. To the best of our knowledge,
considerations of GPE-based models in similar 3D settings have not been
reported before.

\section{Analytical considerations}

\subsection{The framework for the analysis of the Gross-Pitaevskii equation}

It is relevant to explicitly write the underlying 3D equation (\ref{GPE}) in
the spherical coordinates:
\begin{gather}
i\frac{\partial \psi }{\partial t}=-\frac{1}{2}\left[ \frac{\partial ^{2}}{%
\partial r^{2}}+\frac{2}{r}\frac{\partial }{\partial r}\right.   \notag \\
\left. +\frac{1}{r^{2}}\left( \frac{\partial ^{2}}{\partial \theta ^{2}}%
+\cot \theta \cdot \frac{\partial }{\partial \theta }+U_{0}+\frac{1}{\sin
^{2}\theta }\frac{\partial ^{2}}{\partial \varphi ^{2}}\right) \right] \psi
+\left\vert \psi \right\vert ^{2}\psi .  \label{GPE-single}
\end{gather}%
Stationary solutions to Eq. (\ref{GPE-single}), with chemical potential $\mu
<0$ and vorticity which is represented by the magnetic quantum number $m$,
are looked for as
\begin{equation}
\psi =\exp \left( -i\mu t+im\varphi \right) u^{(m)}\left( r,\theta \right) ,
\label{psi-um}
\end{equation}%
cf. Eq. (\ref{psi-u}), with real function $u^{(m)}\left( r,\theta \right) $
satisfying the stationary equation,%
\begin{gather}
\mu u^{(m)}=-\frac{1}{2}\left[ \frac{\partial ^{2}}{\partial r^{2}}+\frac{2}{%
r}\frac{\partial }{\partial r}\right.   \notag \\
\left. +\frac{1}{r^{2}}\left( \frac{\partial ^{2}}{\partial \theta ^{2}}%
+\cot \theta \cdot \frac{\partial }{\partial \theta }+\left( U_{0}-\frac{%
m^{2}}{\sin ^{2}\theta }\right) \right) \right] u^{(m)}+\left(
u^{(m)}\right) ^{3}.  \label{u2}
\end{gather}

To eliminate the singularity, $\sim 1/r$, of the solution, we perform
substitution (\ref{uv}), i.e.,%
\begin{equation}
u^{(m)}\left( r,\theta \right) \equiv r^{-1}V^{(m)}\left( r,\theta \right) ,
\label{uv2}
\end{equation}%
in Eq. (\ref{u2}), which leads to an equation for $V^{(m)}\left( r,\theta
\right) $ [cf. Eq. (\ref{chi})],%
\begin{equation}
2\mu r^{2}V^{(m)}=-r^{2}\frac{\partial ^{2}V^{(m)}}{\partial r^{2}}-\left(
\frac{\partial ^{2}}{\partial \theta ^{2}}+\cot \theta \cdot \frac{\partial
}{\partial \theta }+\left( U_{0}-\frac{m^{2}}{\sin ^{2}\theta }\right)
\right) V^{(m)}+2\left( V^{(m)}\right) ^{3}.  \label{V}
\end{equation}%
For the consideration of the form of eigenmodes in the limit of $%
r\rightarrow 0$, one may define, straightforwardly,%
\begin{equation}
V^{(m)}\left( r=0,\theta \right) \equiv v^{(m)}(\theta ),  \label{Vv}
\end{equation}%
and directly set $r=0$ in Eq. (\ref{V}), which leads, without any
approximation, to the ordinary differential equation with respect to the
angular coordinate:
\begin{equation}
\frac{d^{2}v^{(m)}}{d\theta ^{2}}+\cot \theta \cdot \frac{dv^{(m)}}{d\theta }%
+\left( U_{0}-\frac{m^{2}}{\sin ^{2}\theta }\right) v^{(m)}=2\left(
v^{(m)}\right) ^{3}.  \label{v}
\end{equation}%
Equation (\ref{v}) should be solved in the natural region of $0\leq \theta
\leq \pi $. In fact, Eq. (\ref{Vv}) implies that relevant solutions to Eq. (%
\ref{v}) provide the boundary condition for solutions to Eq. (\ref{V}) at $%
r\rightarrow 0$. The boundary condition at $r\rightarrow \infty $ amounts to
the exponential decay of the wave function:
\begin{equation}
V^{(m)}\left( r\rightarrow \infty ,\theta \right) \sim \exp \left( -\sqrt{%
-2\mu }r\right)  \label{V2}
\end{equation}%
(recall the bound states may exist solely with $\mu <0$).

The presence of the artificial singularity in Eq. (\ref{v}) at $\theta =0$
and $\pi $ suggests a possibility of the existence of singular solutions
with the asymptotic form%
\begin{equation}
v_{\text{\textrm{sing}}}^{(m)}(\theta )\approx v_{0}^{(m)}\theta ^{-1},%
\mathrm{or}~v_{\text{\textrm{sing}}}^{(m)}(\theta )\approx v_{0}^{(m)}(\pi
-\theta )^{-1}.  \label{sing}
\end{equation}%
The substitution of this ansatz in Eq. (\ref{v}) yields%
\begin{equation}
\left( v_{0}^{(m)}\right) ^{2}=\frac{1}{2}\left( 1-m^{2}\right) .
\label{1-m^2}
\end{equation}%
Obviously, it follows from Eq. (\ref{1-m^2}) that the singular solution may
exist only for $m=0$ (in particular, for $m=0$ and $U_{0}=0$ Eq. (\ref{v})
has an exact singular solution, $v^{(0)}\left( U_{0}=0\right) =\left( \sqrt{2%
}\sin \theta \right) ^{-1}$). Actually, the singular solution (\ref{sing})
is irrelevant, as the respective normalization integral includes a divergent
factor, $\int_{0}^{\pi }\left[ v_{\text{\textrm{sing}}}^{(m=0)}(\theta )%
\right] ^{2}\sin \theta d\theta =\infty $.

\subsection{Identifying vortex (sectoral) states with equal magnetic and
orbital quantum numbers, $m=l$}

Vortex-state solutions of Eq. (\ref{V}) with $m=l$, also known as \textit{%
sectoral} ones, are determined by Eq. (\ref{v}) with boundary conditions at $%
\theta =0$ and $\pi $

\begin{equation}
v^{(m)}\sim \theta ^{m},v\sim (\pi -\theta )^{m}  \label{S}
\end{equation}%
(by definition, $m$ does not take negative integer values). Equation (\ref{v}%
) with these boundary conditions admits an exact solution, suggested by the
form of spherical harmonic $\mathrm{Y}_{ll}$, in the form of%
\begin{equation}
v^{(m=l)}(\theta )=v_{0}^{(m=l)}\left( \sin \theta \right) ^{m},  \label{v0}
\end{equation}%
with an infinitesimal amplitude $v_{0}^{(m)}$, at the \textit{threshold value%
} of the strength of the pulling potential,%
\begin{equation}
\left( U_{0}^{(m=l)}\right) _{\mathrm{thr}}=m(m+1)\equiv l(l+1),
\label{eigen}
\end{equation}%
which is an example of the general threshold value (\ref{thr}). Above the
threshold, i.e., at%
\begin{equation}
U_{0}>m(m+1),  \label{existence}
\end{equation}%
amplitude $v_{0}^{(m)}$ in Eq. (\ref{v0}) grows from infinitesimal to finite
values. This fact has a simple explanation: because the orbital kinetic
energy of the state with quantum numbers $l=m$ is
\begin{equation}
E_{\mathrm{kin}}=m(m+1)/\left( 2r^{2}\right) ,  \label{eigenE}
\end{equation}%
condition (\ref{existence}) implies that the energy of pulling to the center
exceeds the kinetic (centrifugal) energy.

In the simplest case, $m=l=1$ (the \textit{p}-wave sectoral state, in terms
of atomic physics), one can substitute expression (\ref{v0}),%
\begin{equation}
v^{(m=l=1)}(\theta )=v_{0}^{(m=l=1)}\sin \theta ,  \label{1}
\end{equation}%
as an approximation, in the nonlinear term of Eq. (\ref{v}), and make use of
identity
\begin{equation}
\left( \sin \theta \right) ^{3}=\frac{1}{4}\left[ 3\sin \theta -\sin
(3\theta )\right] .  \label{sin^3}
\end{equation}%
Neglecting, as usual, the third harmonic in Eq. (\ref{sin^3}) and keeping
the term $\sim \sin \theta $, for small values of the distance from the
threshold, \textit{viz}.,
\begin{equation}
U_{0}-\left( U_{0}^{(m=l=1)}\right) _{\mathrm{thr}}\equiv U_{0}-2\ll 1,
\label{<<}
\end{equation}%
the so approximated Eq. (\ref{v}) predicts the squared amplitude of the
approximate solution (\ref{v0}) as
\begin{equation}
\left( v_{0}^{(m=l=1)}\right) ^{2}\approx \frac{2}{3}\left( U_{0}-2\right) .
\label{S=1}
\end{equation}

Similarly, for the vortex (sectoral) state with $m=l=2$ (the \textit{d}-wave
orbital, in terms of atomic physics), the exact solution (\ref{v0}) with an
infinitesimal amplitude $v_{0}^{(m=l=2)}$,%
\begin{equation}
v^{(m=l=2)}(\theta )=v_{0}^{(m=l=2)}\sin ^{2}\theta ,  \label{2}
\end{equation}%
exists at $U_{0}=6$, as given by Eq. (\ref{eigen}) with $l=2$. To predict
the amplitude of this state above the threshold, one can use the following
approximation:%
\begin{equation}
\sin ^{6}\theta \equiv \frac{15}{16}\sin ^{2}\theta -\frac{3}{8}\sin
^{2}(2\theta )+\frac{1}{16}\sin ^{2}(3\theta )\approx \frac{15}{16}\sin
^{2}\theta ,  \label{6}
\end{equation}%
omitting the second and third harmonics. The substitution of ansatz (\ref{2}%
) and approximation (\ref{6}) in the nonlinear term of Eq. (\ref{v}) with $%
m=2$ produces the squared amplitude of the solution,%
\begin{equation}
\left( v_{0}^{(m=l=2)}\right) ^{2}\approx \frac{8}{15}\left( U_{0}-6\right) ,
\label{S=2}
\end{equation}%
in the case of
\begin{equation}
U_{0}-U_{0}^{(m=l=2)}\equiv U_{0}-6\ll 1.  \label{2<<}
\end{equation}

\subsection{Identifying the states with $m<l$}

For the case of $l=1$ (the \textit{p} state, in terms of atomic orbitals)
and $m=0$, Eq. (\ref{v}) produces a \textit{zonal} \textit{mode}, which is
shaped as a dipole along the $z$ direction. In the limit case of the
infinitesimal amplitude, an exact solution of this type, suggested by the
form of the $\mathrm{Y}_{10}$ spherical harmonic,%
\begin{equation}
v^{(l=1,m=0)}(\theta )=v_{0}^{(l=1,m=0)}\cos \theta ,  \label{cos}
\end{equation}%
exists at the same threshold value, $U_{0}^{(l=1,m=0)}=2$, which is given by
Eq. (\ref{eigen}) for $l=1$. Above the threshold, \textit{viz}., at small $%
U_{0}-2$, the squared amplitude of the mode (\ref{cos}) is%
\begin{equation}
\left( v_{0}^{(l=1,m=0)}\right) ^{2}\approx \frac{2}{3}\left( U_{0}-2\right)
.  \label{dip}
\end{equation}%
It coincides with the one given Eq. (\ref{S=1}) for $l=m=1$, which also
represents the \textit{p}-wave orbital.

Further, in the case of $l=2$ (the \textit{d}-wave state, in terms of atomic
orbitals) and $m=1$, an appropriate ansatz for the respective \textit{%
dipole-vortex} state (also known as a \textit{tesseral} \textit{mode}) is
suggested by the form of spherical harmonic $\mathrm{Y}_{21}$:%
\begin{equation}
v^{(l=2,m=1)}(\theta )=v_{0}^{(l=2,m=1)}\sin \left( 2\theta \right) .
\label{sin2}
\end{equation}%
For $l=2$, the respective threshold value of the strength of the pull to the
center is $U_{0}^{(l=2,m=0)}=6$, as given, once again, by Eq. (\ref{eigen})
with $l=2$. Above the threshold, i.e., at $0<U_{0}-6\ll 1$ (cf. Eq. (\ref%
{2<<})), the squared amplitude of solution (\ref{sin2}) is predicted as
\begin{equation}
\left( v_{0}^{(l=2,m=1)}\right) ^{2}\approx \frac{2}{3}\left( U_{0}-6\right)
,  \label{quadr}
\end{equation}%
cf. Eq. (\ref{S=2}).

For the set of quantum numbers $l=2$ and $m=0$, the \textquotedblleft
sandwich-shaped" ansatz (with structure $\left( -+-\right) $ along the $z$
axis, alias the zonal mode)), is suggested by the corresponding spherical
harmonic $\mathrm{Y}_{20}$, \textit{viz}.,%
\begin{equation}
v^{(l=2,m=0)}(\theta )=v_{0}^{(l=2,m=0)}\left( 1-3\cos ^{2}\theta \right) .
\label{cos^2}
\end{equation}%
This expression with an infinitesimal amplitude gives an exact solution of
Eq. (\ref{v}) at the same threshold, $U_{0}=6$, which corresponds to $l=2$
and $m=2$ or $m=1$, see above. Finally, above the threshold, i.e., at $%
0<U_{0}-6\ll 1$, the approximation similar to that elaborated above predicts
the squared amplitude of solution (\ref{cos^2}) as

\begin{equation}
\left( v_{0}^{(l=2,m=0)}\right) ^{2}\approx \frac{4}{11}\left(
U_{0}-6\right) ,  \label{4/11}
\end{equation}%
cf. Eqs. (\ref{S=2}) and (\ref{quadr}).

Similarly, for $l>2$, the sectoral, tesseral, and zonal states, with $m=l$, $%
0<m<l$, and\ $m=0$, respectively, exist at $U_{0}>l(l+1)$. Above the
existence threshold, amplitudes of these states scale as $\sqrt{U_{0}-l(l+1)}
$, cf. Eqs. (\ref{S=1}), and (\ref{dip}) for $l=1$, and Eqs. (\ref{S=2}), (%
\ref{quadr}), and (\ref{4/11}) for $l=2$.

\subsection{The global approximation for the states with $l\geq 1$}

Similar to the interpolation approximation for the GS, represented by Eqs. (%
\ref{inter}) and (\ref{mu}), an approximation for the states carrying the
angular momentum can be constructed by juxtaposing expressions (\ref{v0}),
or\ (\ref{2}), or (\ref{cos}), or (\ref{sin2}), or (\ref{cos^2}) for $%
r\rightarrow 0$, and Eq. (\ref{V2}) for $r\rightarrow \infty $ in Eqs. (\ref%
{psi-um}) and (\ref{uv2}):%
\begin{gather}
\psi _{\mathrm{interpol}}=\exp \left( -i\mu t+im\varphi \right)
r^{-1}V^{\left( l,m\right) }\left( r,\theta \right) ,  \notag \\
V^{\left( l,m\right) }\left( r,\theta \right) \approx v^{\left( l,m\right)
}(\theta )\exp \left( -\sqrt{-2\mu }r\right)  \label{exp}
\end{gather}

The family of the bound states is characterized by their norm, considered as
a function of $\mu $, cf. Eq. (\ref{mu}). In terms of the interpolation
approximation (\ref{exp}), the norm is written as
\begin{equation}
N_{\mathrm{interpol}}=\frac{I^{(l,m)}}{\sqrt{-2\mu }},~I^{(l,m)}\equiv \pi
\int_{0}^{\pi }\left[ v^{\left( l,m\right) }(\theta )\right] ^{2}\sin \theta
d\theta .  \label{norm}
\end{equation}%
As well as its GS counterpart (\ref{mu}), this relation satisfies the
anti-Vakhitov-Kolokolov criterion, which is a necessary stability condition
for these bound states.

Comparison of this approximation with numerical results demonstrates that it
provides reasonable accuracy for the vortex (sectoral) states with $m=l$,
while for the tesseral and zonal ones, with $m<l$, a discrepancy with the
numerical solutions is relatively large. For the sectoral states, the
substitution of the angular wave function (\ref{v0}) in Eq. (\ref{norm})
yields the following result:%
\begin{equation}
N_{\mathrm{interpol}}^{(m=l)}(\mu )=\frac{2\pi (2m)!!}{(2m+1)!!\sqrt{-2\mu }}%
\left( v_{0}^{(m=l)}\right) ^{2}.  \label{N(interpol)}
\end{equation}%
%
%
%
%
%
%
In this expression, coefficients $\left( v_{0}^{(m=l)}\right) ^{2}$ should
be substituted as per Eqs. (\ref{S=1}) and (\ref{S=2}).
Note that the scaling
\begin{equation}
N\sim 1/\sqrt{-\mu },  \label{scaling}
\end{equation}%
featured by Eq. \ref{N(interpol)},
is an exact property of solutions generated by Eq. (\ref{u2}), irrespective
of the applicability of any approximation.

\subsection{The Thomas-Fermi (TF) approximation}

In the case of a large strength of the pull-to-the-center potential, $%
U_{0}\gg 1$, the angular equation (\ref{v}) admits the application of the
Thomas-Fermi (TF) approximation, which neglects the derivatives in that
equation, yielding%
\begin{equation}
\left( v_{\mathrm{TF}}^{(m)}(\theta )\right) ^{2}=\left\{
\begin{array}{c}
\frac{1}{2}\left( U_{0}-\frac{m^{2}}{\sin ^{2}\theta }\right) ,~\mathrm{at}%
~\arcsin \left( \frac{m}{\sqrt{U_{0}}}\right) <\theta <\pi -\arcsin \left(
\frac{m}{\sqrt{U_{0}}}\right) , \\
0,~\mathrm{at}~0\leq \theta <\arcsin \left( \frac{m}{\sqrt{U_{0}}}\right) ~~%
\mathrm{and}~~\pi -\arcsin \left( \frac{m}{\sqrt{U_{0}}}\right) <\theta \leq
\pi .%
\end{array}%
\right.  \label{TF}
\end{equation}%
The TF approximation is relevant in the case of $U_{0}>m^{2}$, when $\arcsin
\left( m/\sqrt{U_{0}}\right) $ exists. As any TF approximation, expression (%
\ref{TF}) is continuous, featuring discontinuities of the first derivative, $%
dv/d\theta $, at points $\theta =\arcsin \left( m/\sqrt{U_{0}}\right) $ and $%
\theta =\pi -\arcsin \left( m/\sqrt{U_{0}}\right) $. The largest squared
value of the approximate solution (\ref{TF}) is
\begin{equation}
\left( v_{\mathrm{TF}}^{(m)}\left( \theta =\frac{\pi }{2}\right) \right)
^{2}=\frac{1}{2}\left( U_{0}-m^{2}\right) .  \label{TFmax}
\end{equation}

For large $U_{0}$, the TF approximation applies to the full equation (\ref{V}%
) as well: neglecting both the radial and angular derivatives in it, one
obtains%
\begin{equation}
\left( V_{\mathrm{TF}}^{(m)}(r,\theta )\right) ^{2}=\left\{
\begin{array}{c}
\frac{1}{2}\left( U_{0}-\frac{m^{2}}{\sin ^{2}\theta }\right) -|\mu |r^{2},~%
\mathrm{at}~\frac{m^{2}}{2\sin ^{2}\theta }+|\mu |r^{2}<\frac{U_{0}}{2}, \\
0,~\mathrm{at}~\frac{m^{2}}{2\sin ^{2}\theta }+|\mu |r^{2}>\frac{U_{0}}{2}.%
\end{array}%
\right.  \label{TFfull}
\end{equation}%
In this solution, it is again taken into account that $\mu $ is negative for
bound states. As well as the TF approximation (\ref{TF}), the result (\ref%
{TFfull}) is relevant for $U_{0}>m^{2}$.

Further, if the TF approximation is used at $r\rightarrow 0$, in the form of
Eq. (\ref{TF}), but not globally, the corresponding interpolation-type
approximation, similar to Eq. (\ref{exp}), is%
\begin{equation}
\left( u_{\mathrm{interpom}}^{(m)}\left( r,\theta \right) \right) _{\mathrm{%
TF}}=\frac{v_{\mathrm{TF}}^{(m)}(\theta )}{r}\exp \left( -\sqrt{-2\mu }%
r\right).  \label{interpol-TF}
\end{equation}

\section{Numerical results}

To produce stationary wave functions of the bound states in the numerical
form, it is relevant, first, to substitute ansatz (\ref{psi-um}) in the
underlying GPE\ (\ref{GPE-single}), where $u^{(m)}$ is written as per Eq. (%
\ref{uv2}), but with time-dependent $V^{(m)}\left( r,\theta ,t\right) $. The
substitution leads to a time-dependent version of Eq. (\ref{V}), \textit{viz}%
.,
\begin{equation}
2ir^{2}\frac{\partial V^{(m)}}{\partial t}=-r^{2}\frac{\partial ^{2}V^{(m)}}{%
\partial r^{2}}-\left[ \frac{\partial ^{2}}{\partial \theta ^{2}}+\cot
\theta \cdot \frac{\partial }{\partial \theta }+\left( U_{0}-\frac{m^{2}}{%
\sin ^{2}\theta }\right) \right] V^{(m)}+2\left\vert V^{(m)}\right\vert
^{2}V^{(m)}.  \label{V(t)}
\end{equation}%
The use of the reduced wave function $V^{(m)}\left( r,\theta \right) $ makes
it possible to present the results free from the singular factor $r^{-1}$
(see Eq. (\ref{psi-um})).

Stationary solutions of Eq. (\ref{V}) were found by means of the
imaginary-time simulations \cite{im-time} of Eq. (\ref{V(t)}), performed
with the help of the split-step Fourier algorithm, using a set of $%
1024\times 128$ spatial modes in the $\left( r,\theta \right) $ domain. The
above-mention scaling invariance of Eq. (\ref{u2}) suggest that it is
sufficient to generate the results for a fixed norm, which we set here to be
$N\equiv 1$, while the chemical potential $\mu $ is adjusted to keep this
value of the norm, as per Eq. (\ref{scaling}).

\subsection{The stationary vortex (sectoral) modes with $m=l$}

\begin{figure}[t]
\begin{center}
\includegraphics[height=5.cm]{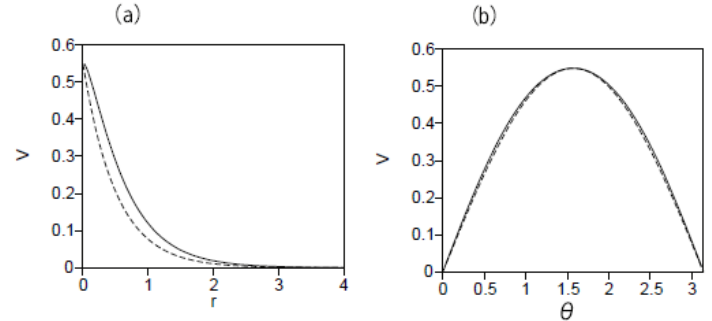}
\end{center}
\caption{(a) The value of the wave function $V^{(m=l=1)}(r)$ (the one in
which the singular factor $r^{-1}$ is removed, see Eqs. (\protect\ref{psi-um}%
) and (\protect\ref{uv2})) at $\protect\theta =\protect\pi /2$, as obtained
from the numerical solution of Eq. (\protect\ref{V}) (the solid line), and
as produced by the exponential fit to it, see Eq. (\protect\ref{0.549}) (the
dashed line). (b) The solid line: the numerically produced profile of $%
v^{(1,1)}(\protect\theta )\equiv V^{(1,1)}\left( r=0,\protect\theta \right) $
(see Eq. (\protect\ref{Vv})), for $U_{0}=2.5$. The chemical potential of
this bound state is $\protect\mu =-1.64$. The dashed line shows the
analytical prediction for the same profile, produced by Eq. (\protect\ref{1}%
), with coefficient $v_{0}^{(m=l=1)}$ replaced by the same fit value, $0.549$%
, as in panel (a) (see Eq. (\protect\ref{0.549})).}
\label{fig1}
\end{figure}

\begin{figure}[t]
\begin{center}
\includegraphics[height=5.cm]{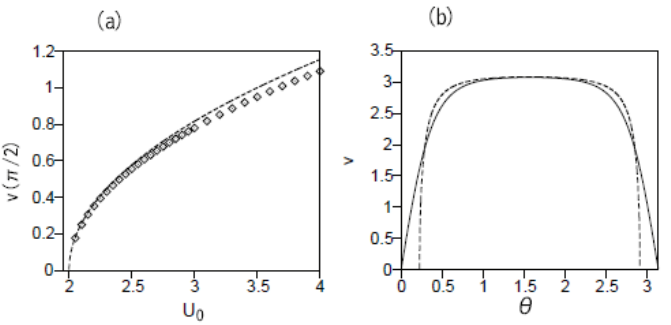}
\end{center}
\caption{(a) The solid curve shows the maximum value of $v^{(1,1)}(\protect%
\theta )\equiv V^{\left( 1,1\right) }\left( r=0,\protect\theta \right) $
(see Eq. (\protect\ref{Vv})), at $\protect\theta =\protect\pi /2$, vs.
strength $U_{0}$ of the pull-to-the-center potential. The dashed curve shows
the analytical prediction for this dependence, as given by Eq. (\protect\ref%
{S=1}). (b) The solid curve shows the numerically produced profile of $%
v^{(1)}(\protect\theta )$ for Eq.~(\protect\ref{v}) at $U_{0}=20,$ while the
dashed one is its counterpart predicted by the TF approximation (\protect\ref%
{TF}).}
\label{fig2}
\end{figure}

First, Figs. \ref{fig1} and \ref{fig2} display essential results for the
vortex (sectoral) bound state with quantum numbers $m=l=1$, and compares
them to the analytical predictions reported in the previous section. For the
strength of the pulling potential $U_{0}=2.5$, the chemical potential of the
bound state corresponding to the fixed norm, $N=1$, which is presented in
Fig. \ref{fig1}, is $\mu =-1.64$. In this case, the $r$-dependence of the
numerically found value of $V^{(m=l=1)}\left( r,\theta =\pi /2\right) $,
which is the maximum of wave function $V^{(m=l=1)}\left( r,\theta \right) $
for fixed $r$, is presented by the solid line in Fig. \ref{fig1}(a). The
dashed line in the same panel displays an exponential fit to the numerical
result, \textit{viz}.,%
\begin{equation}
V^{(m=l=1)}\left( r,\theta =\pi /2\right) \approx 0.549\exp \left( -\sqrt{%
-2\mu }r\right) .  \label{0.549}
\end{equation}%
The respective interpolation approximation, produced by Eqs. (\ref{1}), (\ref%
{S=1}), and (\ref{exp}), is%
\begin{equation}
V^{(m=l=1)}\left( r,\theta =\pi /2\right) \approx \sqrt{\frac{2}{3}\left(
U_{0}-2\right) }\exp \left( -\sqrt{-2\mu }r\right) \approx 0.577\exp \left( -%
\sqrt{-2\mu }r\right) ,  \label{0.577}
\end{equation}%
being quite close to its numerically found counterpart (\ref{0.549}).
Further, the numerically obtained profile of $v^{(1,1)}(\theta )\equiv
V^{(1,1)}\left( r=0,\theta \right) $ for the same $U_{0}=2.5$ is displayed
by the solid curve in Fig. \ref{fig1}(b), while the dashed one shows the
corresponding analytical approximation suggested by Eq. (\ref{1}), with the
same coefficient as used for the fitting curve in panel \ref{fig1}(a) (see
Eq. (\ref{0.549})), \textit{viz}., $v^{(1.1)}(\theta )\approx 0.549\sin
\theta $. It is seen that the semi-analytical fit profile is virtually
identical to the numerical one. If the fit coefficient $0.549$, taken from
Eq. (\ref{0.549}), is replaced by its analytically predicted counterpart, $%
\sqrt{(2/3)\left( U_{0}-2\right) }\approx 0.577$ [see Eq. (\ref{0.577})],
the agreement between the numerical findings and analytical predictions
remains very accurate.

The family of the bound states with $m=l=1$ is characterized in Fig. \ref%
{fig2}(a) by the numerically found dependence of the largest value of their
wave function, $V^{(1,1)}\left( r=0,\theta =\pi /2\right) \equiv
v^{(m=l=1)}(\theta =\pi /2)$, on the potential strength $U_{0}$, plotted
along with the analytical prediction (\ref{S=1}). Although the prediction
was derived under the condition of $0<U_{0}-2\ll 1$, it is seen that, even
for $U_{0}=4$, the analytical result remains a highly accurate one.

The adequacy of the TF approximation for large values of the potential's
strength $U_{0}$ is corroborated by Fig. \ref{fig2}(b), which displays a
numerically generated profile of $v^{(1)}(\theta )$ for $U_{0}=20$, along
with its TF-predicted counterpart, given by Eq. (\ref{TF}). It is observed
that the approximation is reasonable. In particular, the largest value of
the wave function, given by expression (\ref{TFmax}), is practically
identical to its numerically obtained counterpart.

\begin{figure}[t]
\begin{center}
\includegraphics[height=5.cm]{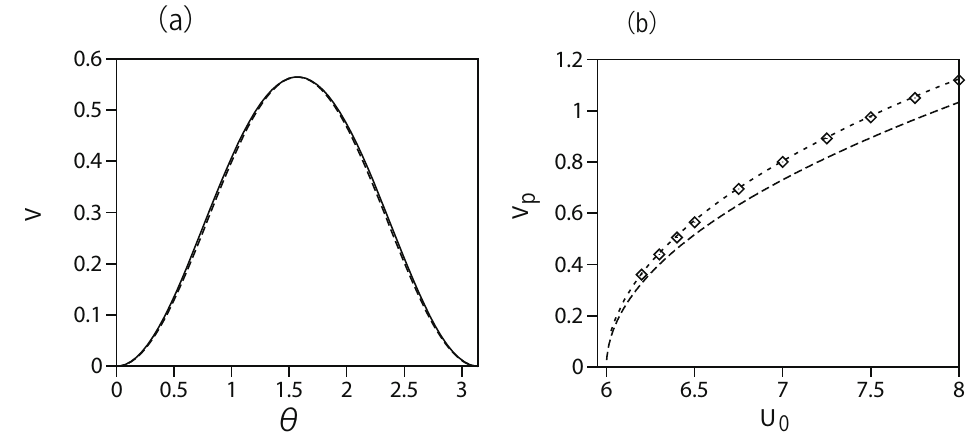}
\end{center}
\caption{(a) The solid line: the numerically found profile of the vortex
(sectoral) state, $v^{(2,2)}(\protect\theta )\equiv V^{(2,2)}\left( r=0,%
\protect\theta \right) $ (see Eq. (\protect\ref{Vv})), for $U_{0}=6.5$. The
chemical potential of this bound state is $\protect\mu =-1.44$. The
analytical prediction for the same profile, as produced by Eq. (\protect\ref%
{2}), is shown by the dashed line. (b) The solid curve: the maximum value of
$v^{(2,2)}(\protect\theta )\equiv V^{\left( 2,2\right) }\left( r=0,\protect%
\theta \right) $ (see Eq. (\protect\ref{Vv})), at $\protect\theta =\protect%
\pi /2$, vs. the potential's strength $U_{0}$. The dashed curve: the
analytical prediction for the same dependence, as given by Eq. (\protect\ref%
{S=2}).}
\label{fig4}
\end{figure}

For the higher-order vortex (sectoral) mode with $m=l=2$, an example of the
numerically found profile, $V^{(2,2)}\left( r=0,\theta \right) \equiv
v^{(2,2)}(\theta )$, is displayed in Fig. \ref{fig4}(a) for $U_{0}=6.5$,
along with the analytically predicted profile, $0.564\sin ^{2}\theta $ (see
Eq. (\ref{S=2})), with the coefficient produced by the fit procedure. The
chemical potential of this state is $\mu =-1.44$. The analytical prediction
is reproduced by the numerical solution practically exactly.

The family of the vortex modes with $m=l=2$ is represented in Fig. \ref{fig4}%
(b) by the dependence of $V^{(m=l=2)}\left( r=0,\theta =\pi /2\right) \equiv
v^{(m=l=2)}(\theta =\pi /2)$ on the potential's strength $U_{0}$, as found
from the numerical solution and predicted by Eq. (\ref{S=2}). In comparison
with the similar dependences shown in Fig. \ref{fig2}(a) for $m=l=1$, the
discrepancy between the numerical and analytical results is somewhat larger,
but the analytical approximation is still quite reasonable even for $U_{0}=8$%
, when the formal applicability condition (\ref{2<<}) does not hold.

\subsection{The bound states with $m<l$}

\begin{figure}[t]
\begin{center}
\includegraphics[height=5.cm]{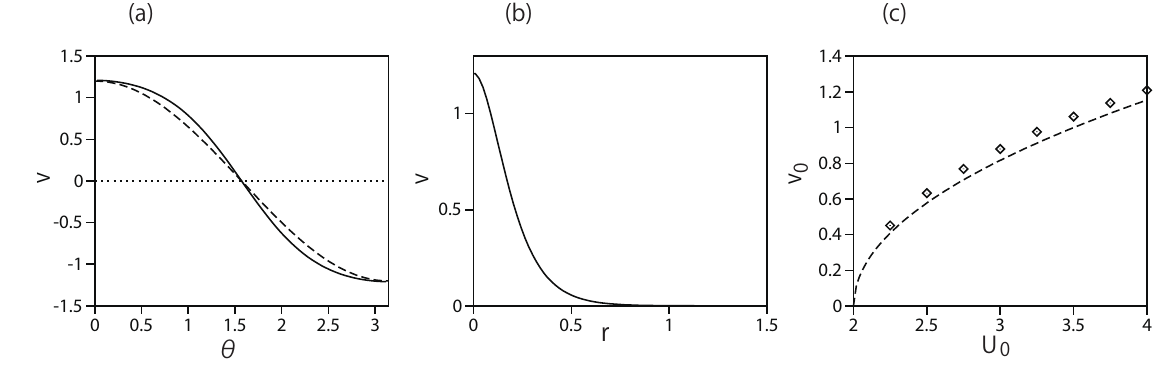}
\end{center}
\caption{(a) The solid line: the numerically found profile of the
dipole-shaped (zonal) state with $l=1$, $m=0$, i.e.. $v^{(1,0)}(\protect%
\theta )\equiv V^{(1,0)}\left( r=0,\protect\theta \right) $ (see Eq. (%
\protect\ref{Vv})), for $U_{0}=4$. The chemical potential of this bound
state is $\protect\mu =-36.0$. The analytical prediction for the same
profile, as produced by Eq. (\protect\ref{cos}), is shown by the dashed
line. (b) The radial profile $V(r,\protect\theta =0)$ (the solid line)
produced by the numerical solution. (c) The solid curve: the maximum value
of $v^{(1,0)}(\protect\theta )\equiv V^{(l,0)}\left( r=0,\protect\theta %
\right) $ (see Eq. (\protect\ref{Vv})), at $\protect\theta =\protect\pi /2$,
vs. the potential's strength $U_{0}$. The dashed curve: the analytical
prediction for the same dependence, as given by Eq. (\protect\ref{dip}).}
\label{fig5}
\end{figure}

An example of the numerically found angular profile of the dipole (zonal)
mode, with $l=1$, $m=0$ and $U_{0}=4$, is plotted by the solid curve in Fig. %
\ref{fig5}(a). The dashed line is its analytically predicted counterpart (%
\ref{cos}), with the fit coefficient $v_{0}^{(1,0)}=1.209$, while the
respective analytical expression (\ref{dip}) yields a close value, $%
v_{0}^{(1,0)}\approx 1.155$. The chemical potential of this bound state is $%
\mu =-36$. The radial dependence $V(r,\theta =0)$ is shown in Fig. \ref{fig5}%
(b) (the solid line). The family of the dipole-shaped (zonal) states is
characterized in Fig. \ref{fig5}(c) by dependences of $V^{(l=1,m=0)}\left(
r=0,\theta =\pi /2\right) \equiv v^{l=1,m=0)}(\theta =\pi /2)$ on strength $%
U_{0}$ of the pulling potential, as found from the numerical solution, and
as predicted analytically by Eq. (\ref{S=2}). It is seen from these figures
that the accuracy of the analytical approximation is reasonable, even for
values of $U_{0}-2$ which are not small.

\begin{figure}[t]
\begin{center}
\includegraphics[height=5.cm]{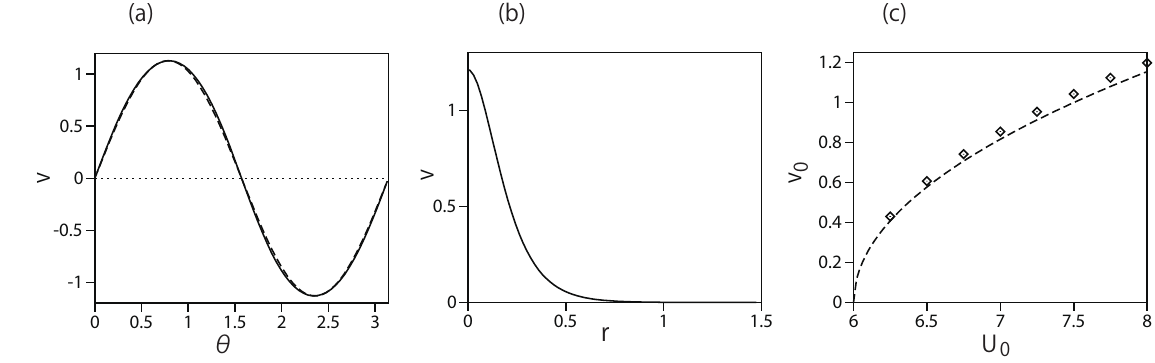}
\end{center}
\caption{(a) The solid line: the numerically found profile of the
dipole-vortex (tesseral) state with $l=2$, $m=1$, i.e.. $v^{(2,1)}(\protect%
\theta )\equiv V^{(2,1)}\left( r=0,\protect\theta \right) $ (see Eq. (%
\protect\ref{Vv})), for $U_{0}=8$. The chemical potential of this bound
state is $\protect\mu =-51$. The analytical form of the same profile is
given by Eq. (\protect\ref{sin2}), as shown by the dashed line. (b) The
solid line is the numerically found radial profile $V(r,\protect\theta =%
\protect\pi /4)$ at $U_{0}=8$. (c) The solid curve: the maximum value of $%
v^{(2,1)}(\protect\theta )\equiv V^{(2,1)}\left( r=0,\protect\theta =\protect%
\pi /2\right) $ (see Eq. (\protect\ref{Vv})) vs. the potential's strength $%
U_{0}$. The dashed curve: the analytical prediction for the same dependence,
as given by Eq. (\protect\ref{quadr}).}
\label{fig6}
\end{figure}

Further, the solid and dashed curves in Fig. \ref{fig6}(a) show,
respectively, an example of the numerically obtained profile of the
vortex-dipole (tesseral) mode, with $l=2$, $m=1$ and $U_{0}=8$, and its
analytical counterpart, produced by Eq. (\ref{sin2}) with the fit
coefficient $v_{0}^{(2,1)}=1.13$ (the chemical potential of this state is $%
\mu =-51$). Note that Eq. (\ref{quadr}) gives the analytical approximation
for the latter coefficient as $v_{0}^{(2,1)}=1.155$, which is very close to
the fit value which makes the numerical and analytical profiles virtually
identical in Fig. \ref{fig6}(a). Figure \ref{fig6}(b) shows the numerically
found radial dependence $V(r,\theta =\pi /4)$ at $U_{0}=8$. The family of
the tesseral states is characterized in Fig. \ref{fig6}(c) by dependences of
$V^{(l=2,m=1)}\left( r=0,\theta =\pi /2\right) \equiv v^{(l=2,m=1)}(\theta
=\pi /2)$ on the potential's strength $U_{0}$, as given by numerical
solution, and as predicted by Eq. (\ref{quadr}).\ It is seen from these
figures that the accuracy of the analytical approximation is reasonable,
even for values of $U_{0}-2$ which are not small.

\begin{figure}[t]
\begin{center}
\includegraphics[height=5.cm]{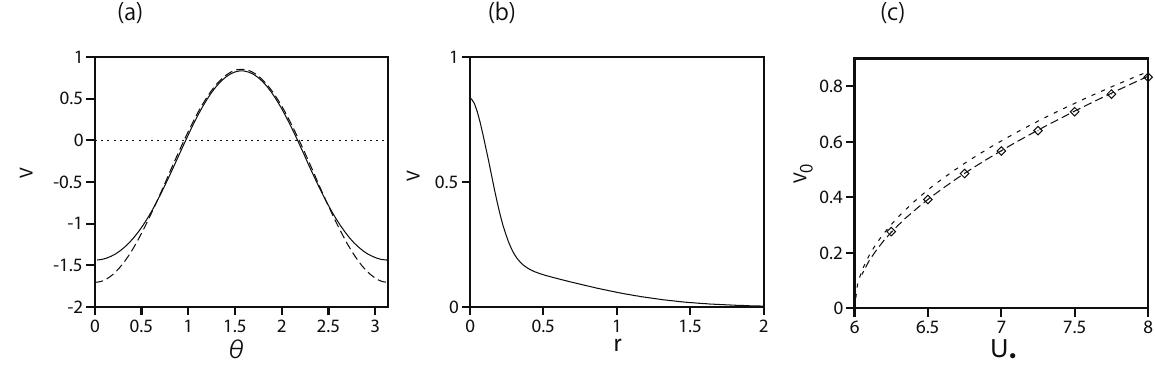}
\end{center}
\caption{(a) The solid line: the numerically found profile of the
\textquotedblleft sandwich-shaped" (zonal) state with $l=2$, $m=0$, i.e., $%
v^{(2,1)}(\protect\theta )\equiv V^{(2,1)}\left( r=0,\protect\theta \right) $
(see Eq. (\protect\ref{Vv})), for $U_{0}=8$. The chemical potential of this
bound state is $\protect\mu =-30.97$. The analytical form of the same
profile is predicted by Eq. (\protect\ref{cos^2}), as shown by the dashed
line. (b) The solid liine: the numerically found radial profile $V(r,\protect%
\theta =\protect\pi /2)$ at $U_{0}=8$. (c) The solid curve: the maximum
value of $v^{(2,0)}(\protect\theta )\equiv V^{(2,0)}\left( r=0,\protect%
\theta =\protect\pi /2\right) $ (see Eq. (\protect\ref{Vv})) vs. the
potential's strength $U_{0}$. The dashed curve: the analytical prediction
for the same dependence, as given by Eq. (\protect\ref{4/11}).}
\label{fig7}
\end{figure}

Finally, numerical results for the \textquotedblleft sandwich-shaped"
(zonal) state with $l=2$, $m=0$ are presented in Fig. \ref{fig7}. An example
of the respective profile of $v^{(l=2,m=0)}(\theta )$ is plotted in Fig. \ref%
{fig7}(a) for $U_{0}=8$ and chemical potential $\mu =-30.97$, along with its
analytically-predicted counterpart (\ref{cos^2}), with coefficient $%
v_{0}^{(l=2,m=0)}=0.852$, which is predicted by Eq. (\ref{4/11}). Figure \ref%
{fig7}(b) shows the numerically found radial dependence $V(r,\theta =\pi /2)$
at $U_{0}=8$. The family of the zonal states is characterized by the
dependence of $V^{(l=2,m=0)}\left( r=0,\theta =\pi /2\right) \equiv
v^{(l=2,m=0)}(\theta =\pi /2)$ on the potential's strength $U_{0}$, which is
plotted in Fig. \ref{fig7}(c), along with the respective analytical
prediction given by Eq. (\ref{4/11}). In this case too, the accuracy of the
analytical approximation is reasonable.

\section{Conclusion}

A solution for the problem of the quantum collapse in the gas of particles
pulled to the center by the well-known potential, $U(r)=-U_{0}/\left(
2r^{2}\right) $, was reported in Ref. \cite{HS1}. Repulsive collisions
between the particles, represented by the cubic term in the respective GPE
(Gross-Pitaevskii equation), suppress the collapse and create the otherwise
missing spherically symmetric GS (ground state), alias the \textit{s}-wave
orbital, in terms of atomic physics. However, the previous analysis did not
address bound states with reduced symmetry, which carry angular momentum.
The present work reports results of the systematic analysis of such states,
which are labeled, as usual, by the orbital and magnetic quantum numbers, $l$
and $m$. These states exist when the strength of the potential exceeds a
threshold value, $U_{0}>l(l+1)$, which does not depend on $m$. At a
relatively small distance from the threshold, the sectoral, tesseral, and
zonal states, with $m=l$, $0<m<l$, and $m=0$, respectively, are found in the
approximate analytical form for $l=1$ and $2$ (alias the \textit{p}- and
\textit{d}-wave orbitals). In the general case, these states are found in
the numerical form, which corroborates the accuracy of the analytical
approximation.

In physical units, characteristics of the predicted stationary states are
similar to the above-mentioned ones for the GS. In particular, in the case
of the gas containing $\sim 10^{5}$ dipolar particles with the dipole moment
$\sim 1$ Debye, pulled to the central electric charge, the eigenstates
carrying the angular momentum will have the size of several microns.
Essentially the same size of the sectoral, tesseral, and zonal eigenstates
is expected in the gas of $\sim 10^{6}$ light atoms pulled to the center by
the converging optical potential defined as per Eqs. (\ref{Udip}) and (\ref%
{I}) with illumination power $\sim 10^{-3}$ W,

The next objective of the analysis is to accurately investigate stability of
the stationary modes, found above in the numerical and approximate
analytical forms. A direct approach to the stability analysis may be based
on real-time simulations of Eq. (\ref{V(t)}). Technically, this is a
challenging problem due to the presence of the artificial singularities in
Eq. (\ref{V(t)}) at $\theta =0$ and $\pi $. Preliminary simulations
demonstrate that, in some cases, apparent artifacts appear in the form of
growing singularities near $\theta =0$ and $\pi $, resembling the
above-mentioned irrelevant stationary singular solutions given by Eqs. (\ref%
{sing}) and (\ref{1-m^2}). Accurate dynamical simulations is a subject for a
separate work, which will be reported elsewhere. In any case, it is relevant
to mention that the well-known splitting instability of vortex states in
nonlinear systems with self-attractive nonlinearity \cite{Michinel,Pego},
which is driven by azimuthal perturbations, has no reason to exist in the
present case of the self-repulsion.

As a further extension of this work, it may also be interesting to analyze
dynamical excitation of the states with angular momentum from the GS by
pulses of an incident vortex field, and to consider Rabi oscillations
between different bound states driven by an ac field. Also relevant is a
possibility to consider the \textit{p}- and \textit{d}-wave orbitals in the
framework of the many-body theory, instead of the mean-field approximation,
following the lines of Ref. \cite{GEA}.

\section*{Funding Information}

The work of B.A.M. is supported, in part, by the Israel Science Foundation
through Grant No. 1695/22.

\section*{Author Contributions}

H.S.: numerical and analytical calculations, production of figures, drafting
the manuscript. B.A.M.: concept of the work, analytical calculations,
drafting the paper.

\section*{Conflict of Interest}

The authors declare no conflicts of interests related to this paper.

\end{document}